\documentclass[epj]{svjour}
\usepackage{epsfig}
\begin{document}
\title{Study of the reactions \boldmath ${\bf \gamma p \rightarrow K^+
\Sigma^{\pm}\pi^{\mp}}$ \unboldmath
at photon energies up to 2.6 GeV\thanks{This work is supported in part
by the Deutsche Forschungsgemeinschaft (SPP KL 980/2-3) and the SFB/TR16}}
\titlerunning{Measurement of the reaction $\gamma p \rightarrow K^ + \Sigma^{\pm}\pi^{\mp}$
at photon energies up to 2.6 GeV}
\author{
I. Schulday\inst{1,3}\thanks{Part of doctoral thesis
 (I.\,Schulday, doctoral thesis,
\newline Bonn University (2004), Bonn-IR-2004-15), http://saphir.
\newline physik.uni-bonn.de/saphir/thesis.html},
R. Lawall\inst{1,5}, J. Barth\inst{1}, K.-H. Glander\inst{1,4}, S.
Goers\inst{1,5}, J. Hannappel\inst{1}, N. J\"open\inst{1}, F.
Klein\inst{1,}\thanks{email: klein@physik.uni-bonn.de},
E.~Klempt\inst{2}, D. Menze\inst{1}, E. Paul\inst{1}, W.J.
Schwille\inst{1} }
\authorrunning{I.\,Schulday et al.}

\institute{
Physikalisches Institut der Universit\"at Bonn, Germany
\and Helmholtz-Institut f\"ur Strahlen- und Kernphysik, Universit\"at Bonn, Germany
\and presently IFB AG, K\"oln, Germany
\and presently TRW Automotive GmbH, Alfdorf, Germany
\and presently T\"UV Nord, Hamburg, Germany
}
\date{Received: date / Revised version: date}

\abstract{ The reactions $\gamma p \rightarrow
K^{+}\Sigma^{\pm}\pi^{\mp}$ were studied with the SAPHIR detector
using a tagged photon beam at the electron stretcher facility ELSA
in Bonn. The decays $\Sigma^{-} \rightarrow n\pi^{-}$ and
$\Sigma^{+} \rightarrow n\pi^{+}, p\pi^0$ were fully reconstructed.
Reaction cross sections were measured as a function of the photon
energy from threshold up to $2.6\,$GeV with considerably improved
statistics compared to a previous bubble chamber measurement. The
cross sections rise monotonously with increasing photon energy. The
two-particle mass distributions of $\Sigma^{\pm}\pi^{\mp}$ and
$K^+\pi^-$ show substantial  production of resonant states.
\PACS{{13.30.-a}{Decays of baryons}   \and
      {14.20.Jn}{hyperons}
} 
} 
\maketitle
\parindent0mm

\section{Introduction}
\label{intro}

Photon-induced reactions on nucleons at low energies are commonly
used to study the excitation of baryonic resonances. A review of
baryon spectroscopy, its aims and its achievements can be found
elsewhere \cite{klempt}. Searches for such resonances  were carried
out in the  SAPHIR experiment analysing the reactions $\gamma p
\rightarrow K^+\Lambda$ \cite{Glan03b}, $\gamma p \rightarrow
K^+\Sigma^0$ \cite{Goers99}, $\gamma p \rightarrow K^0\Sigma^+$
\cite{Lawall05}, $\gamma p \rightarrow \varrho p$ \cite{wu}, $\gamma
p \rightarrow \omega p$ \cite{barthomega}, $\gamma p \rightarrow
\Phi p$~\cite{barthphi}, and $\gamma p \rightarrow \eta^\prime p$
\cite{etaprime}. The measurements presented here  extend this search
to  the  reactions  $\gamma p \rightarrow K^ +
\Sigma^{\pm}\pi^{\mp}$ where  strangeness-zero resonant states might
contribute.

\label{saphir}

The data analysis is  based on 180 million triggered events which
were taken with the magnetic multiparticle  detector SAPHIR
\cite{Schwille} at the $3.5\,$GeV electron stretcher facility ELSA
\cite{Hillert} using a tagged photon beam which covered the photon
energy range from threshold (of the reactions considered here) to
$2.6\,$GeV. A  detailed description of the experiment is given
elsewhere \cite{Glan03b,barthomega}.

The data are available via
internet\footnote{http://saphir.physik.uni-bonn.de/saphir/publications}.

\section{Event reconstruction and event selection}
\label{selection}

\begin{figure}[ht]
\centerline{\includegraphics[width=0.3\textwidth]{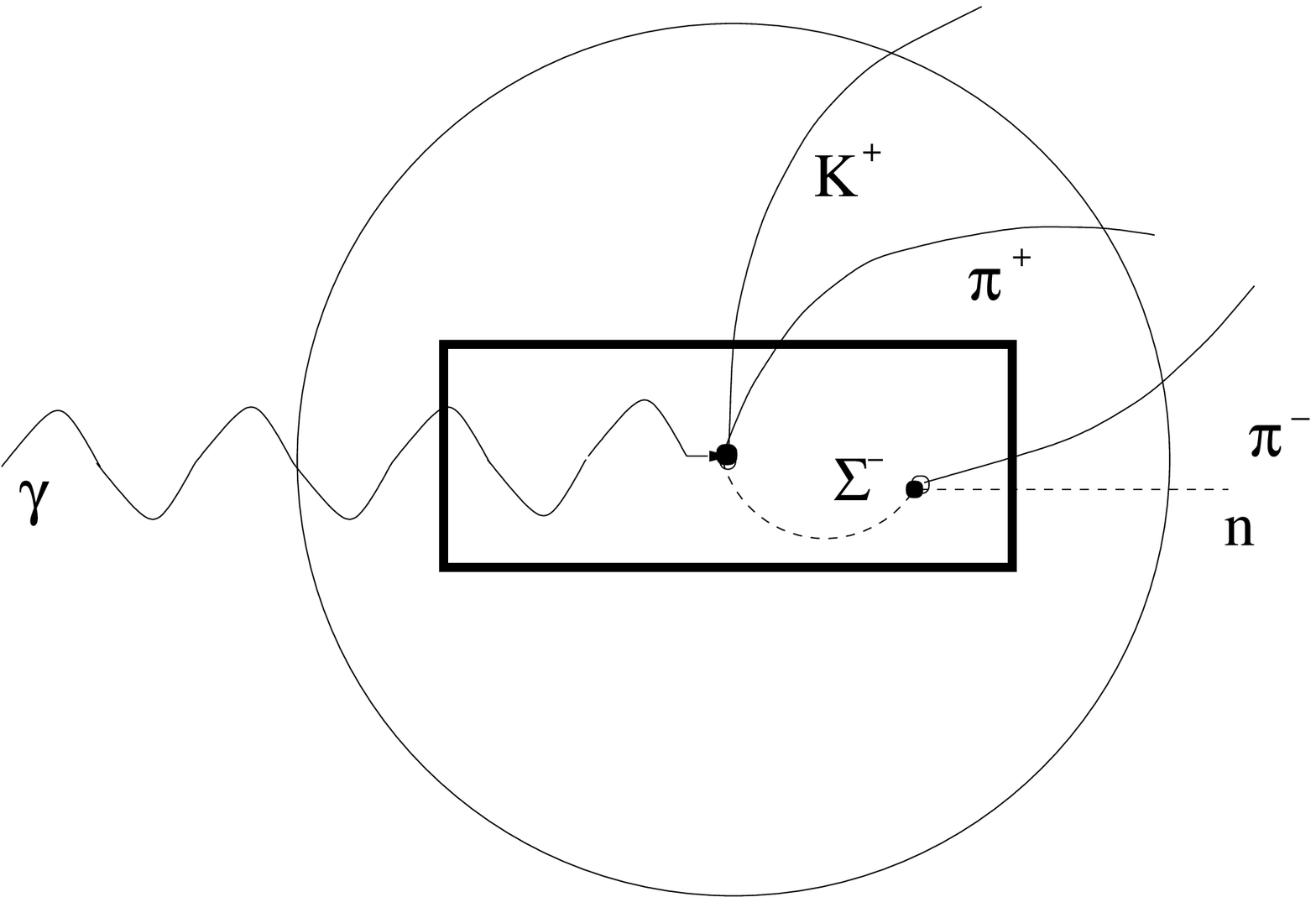}}
\vspace{5mm}
\centerline{\includegraphics[width=0.33\textwidth]{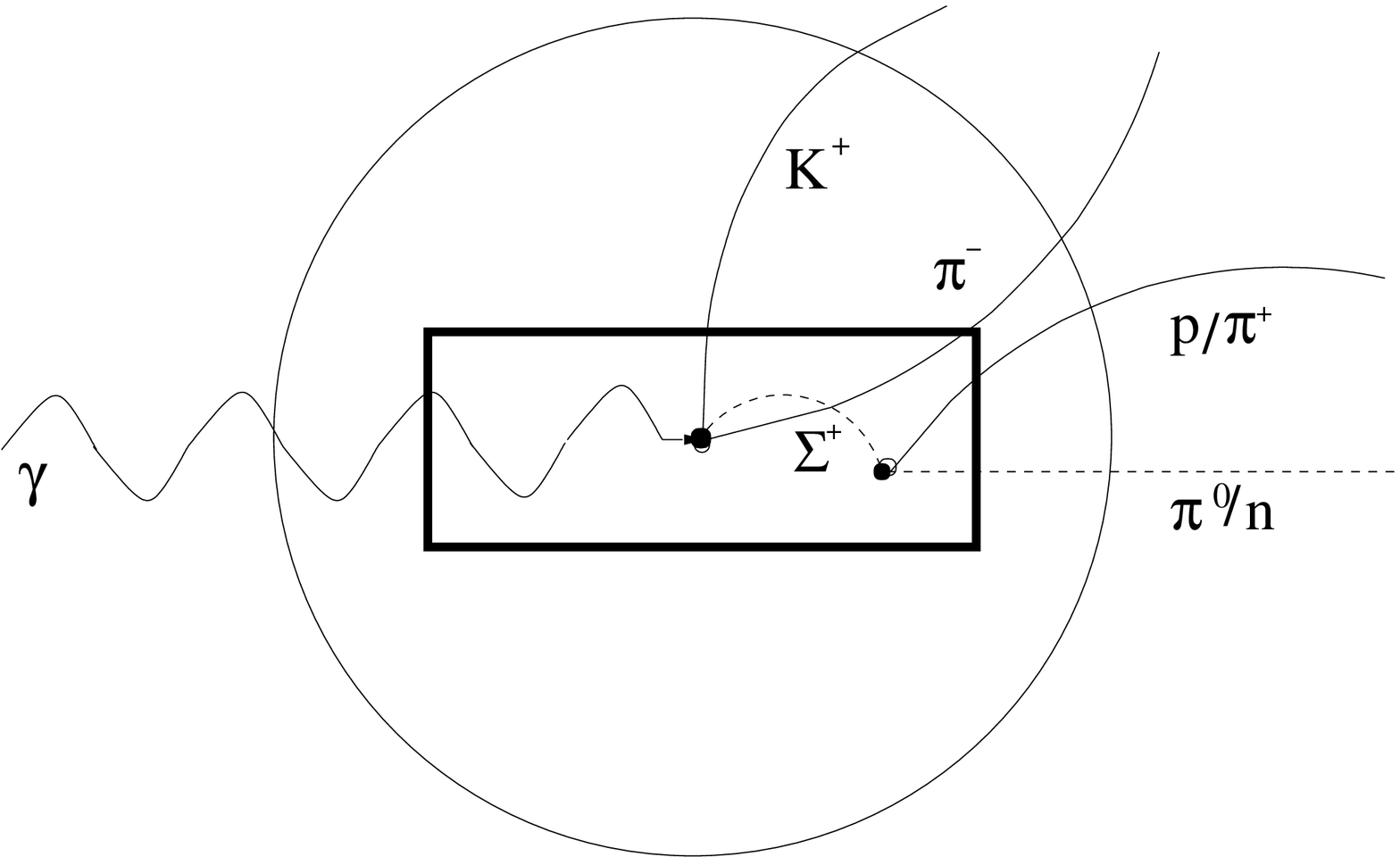}}
\caption{Topologies of the reactions $\gamma p \rightarrow
K^{+}\pi^{+}\Sigma^{-}$ (top) and $\gamma p \rightarrow
K^{+}\pi^{-}\Sigma^{+}$ (bottom) in the target region. Rectangle and
circle represent the target area and the inner layer of the central
drift chamber. The $\Sigma^\pm$ track was not measured.}
\label{pic:saphir2}
\end{figure}

The kinematical reconstruction of the reactions $\gamma p
\rightarrow K^ + \Sigma^{-}\pi^{+}$ with $\Sigma^{-} \rightarrow n
\pi^{-}$, and of $\gamma p \rightarrow K^ + \Sigma^{+}\pi^{-}$ with
either $\Sigma^{+}\rightarrow n \pi^{+}$ or $\Sigma^{+}\rightarrow
p\pi^{0}$, was based on the measurements of the photon energy in the
tagging system and of the three-momenta of the charged particles in
the final states reconstructed in the drift chamber system. The
topology of the events is sketched in fig.\,\ref{pic:saphir2}.

\begin{figure}[ht]
\centerline{
\includegraphics[width=0.5\textwidth]{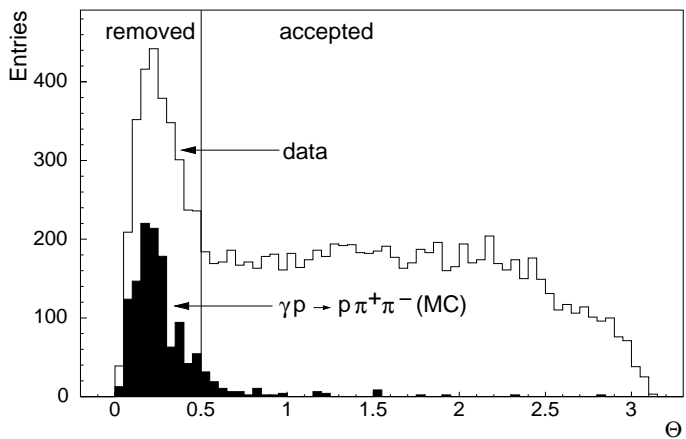}}
\caption{Angular distribution of the decay proton in the
$\Sigma^{+}$ rest system for the decay $\Sigma^{+} \rightarrow
p\pi^{0}$  for data
 and  Monte Carlo simulated background of events from the reaction
$\gamma p \rightarrow p \pi^{+}\pi^{-}$ (black area) which passed
the
 selection cuts. The Monte Carlo event sample was normalised to the photon flux.
 The vertical line indicates the cut which was applied to exclude most
 of this background.}
\label{pic:ppipi}
\end{figure}

In the first step, the primary vertex was searched by combining
pairs out of the three tracks extrapolated into the target region.
The pair with the best matching was accepted. Then the hypotheses
$\gamma p \rightarrow K^ + \Sigma^{\pm}\pi^{\mp}$ were tested by a
kinematical fit which used the photon energy and the  reconstructed
momenta of $K^+$ and $\pi^{\mp}$. The hypothesis with the better fit
probability were tested by a kinematical fit which used the photon
energy and the reconstructed momenta of the particles defining the
primary vertex. The fit determined the 3-momentum of the
$\Sigma^{\pm}$ which allowed us to reconstruct its track downstream
of the accepted primary vertex. In the next step, the  $\Sigma^\pm$
decay vertex was calculated as  intercept of the $\Sigma^{\pm}$
track with the extrapolated track of the third charged particle. The
decay hypotheses $\Sigma^{-} \rightarrow n\pi^{-}$ and $\Sigma^{+}
\rightarrow p\pi^{0}$/$n\pi^{+}$ were tested at the decay vertex by
carrying out corresponding kinematical fits, and the complete
reaction was tested by simultaneous fits at both, the primary and
the decay vertex. Finally, time-of-flight (TOF) measurements carried
out in the range of the geometrical acceptance of the scintillator
hodoscopes were used to reject background from other reactions. For
$\gamma p \rightarrow K^+\Sigma^-\pi^+$ it was required that the
mass assignments, obtained from TOF measurements for the positively
charged particles, had a value below $0.8\,$GeV. This cut removed
events with a proton in the final state. For $\gamma p \rightarrow
K^+\Sigma^+\pi^-$ it was required that the mass assignments were
consistent with the mass values from the fit.

At this stage, the sample of events identified as $\gamma p
\rightarrow K^{+}\Sigma^{+}\pi^{-}$ with $\Sigma^{+}\rightarrow
p\pi^{o}$  still contained substantial background from events due to
the reaction $\gamma p \rightarrow p \pi^{+} \pi^{-}$. The final
states of these reactions have proton and $\pi^{-}$ in common, and
the identification of pions and kaons is not unique because of the
limited time resolution of the time-of-flight (TOF) measurement and
the restriction of geometrical acceptance of the scintillator
hodoscopes.

Figure \ref{pic:ppipi} shows the proton angular distribution in the
$\Sigma^+$ rest system with respect to the momentum of $\Sigma^+$ in
the laboratory system. For comparison, the same distribution of
Monte-Carlo simulated events due to $\gamma p \rightarrow p \pi^{+}
\pi^{-}$ is shown which passed the same selection cuts. The peak in
the data at low angles is qualitatively described  by the simulated
background. An angular cut was applied (vertical line) to remove
most of this background contribution.

\begin{figure}[t!]
\centerline{
\includegraphics[width=0.45\textwidth]{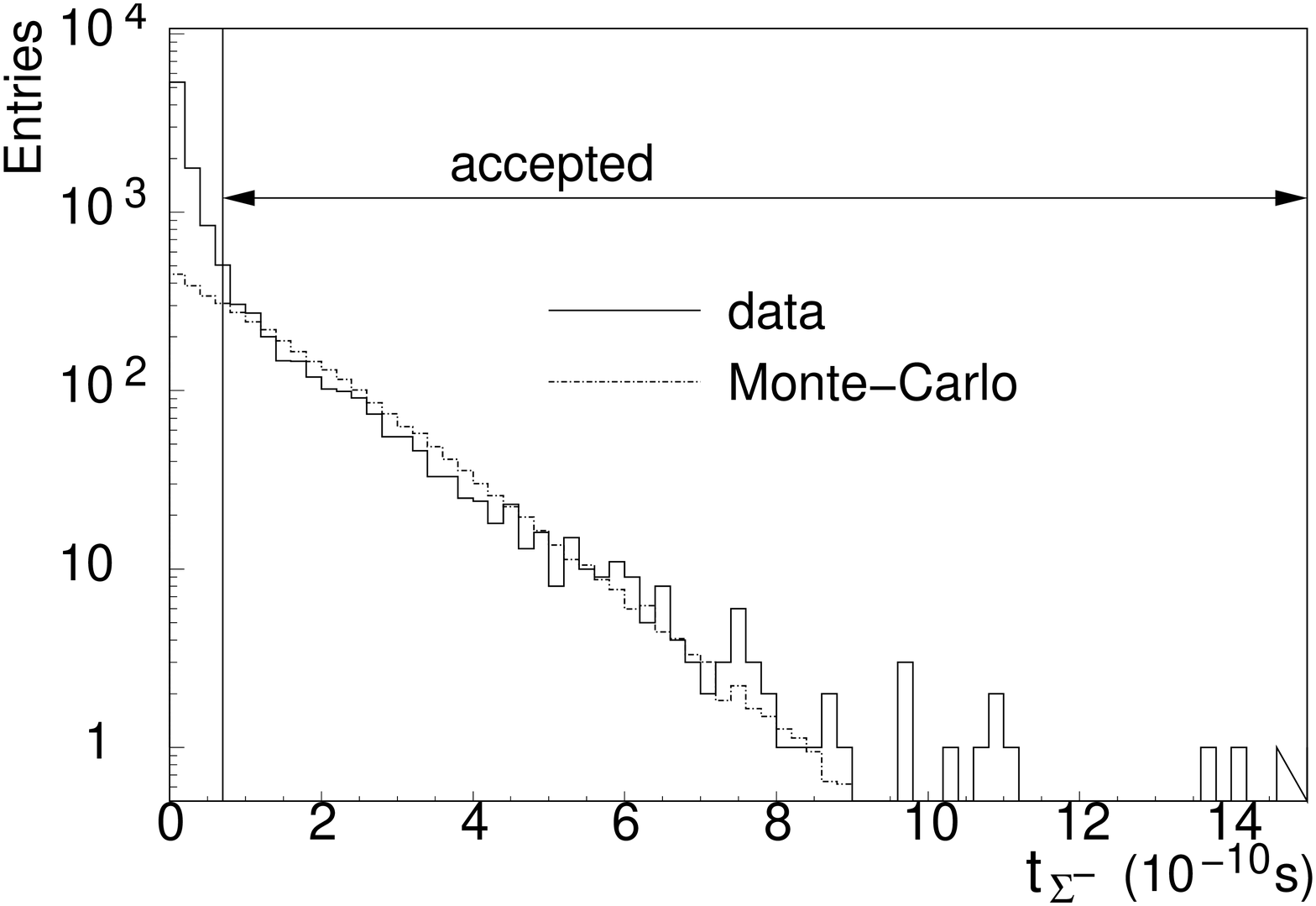}}
\caption{Decay time distribution of the $\Sigma^{-}$ for data (solid
line) and for Monte Carlo simulated events (dashed line), normalised
to the data in the accepted area. The vertical line indicates the
cut applied to the data.}
\label{pic:lebensdauer-}
\centerline{\includegraphics[width=0.45\textwidth]{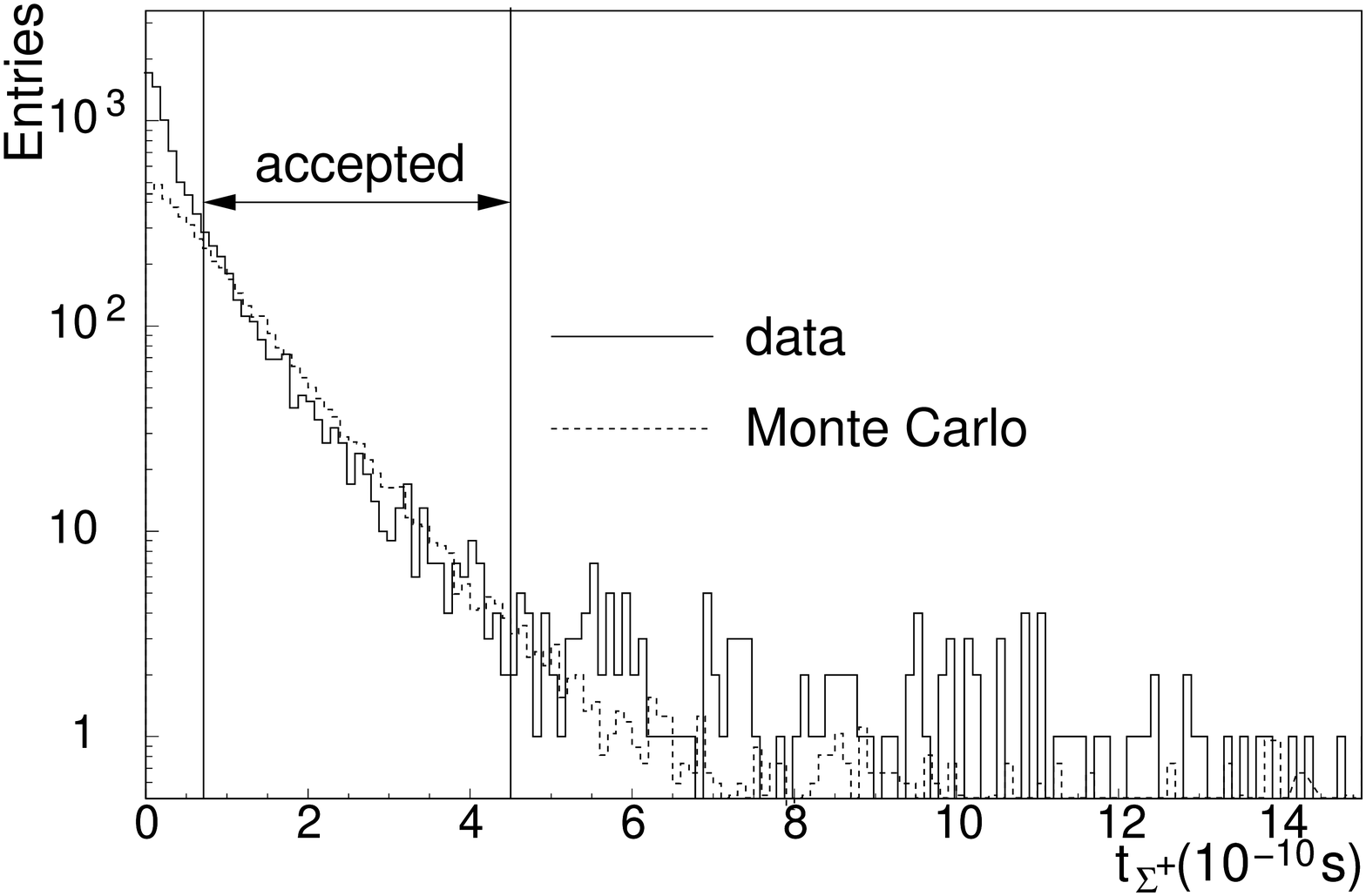}}
\centerline{\includegraphics[width=0.45\textwidth]{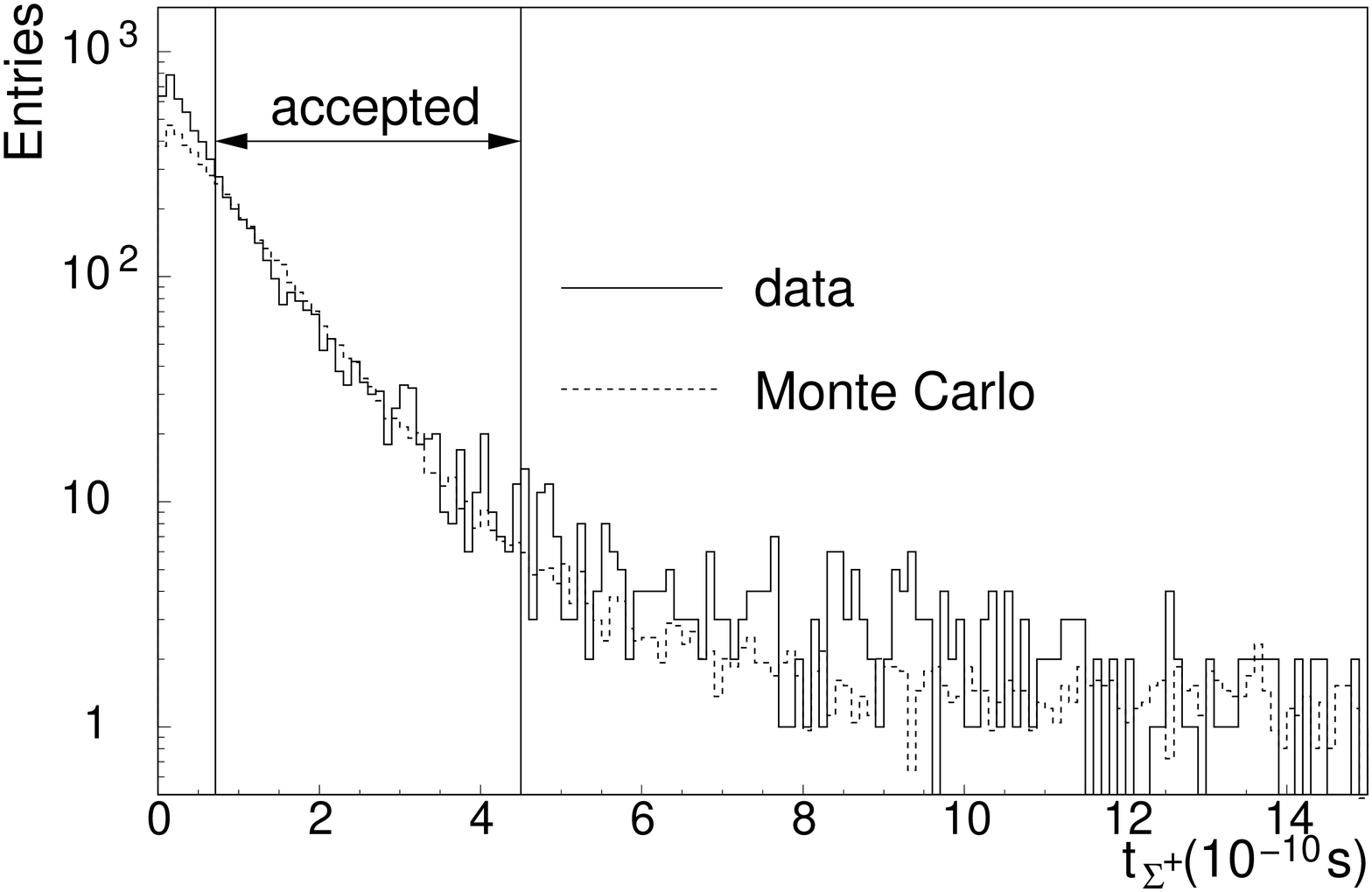}}
\caption{Top: Decay time distribution of the $\Sigma^{+}$ with decay
$\Sigma^{+} \rightarrow n\pi^{+}$ for data (solid line) and Monte
Carlo simulated events (dashed line), normalised to the data in the
accepted area. Bottom: Decay time distribution of the $\Sigma^{+}$
with decay   $\Sigma^{+} \rightarrow p \pi^{0}$ for data (solid
line) and Monte Carlo simulated events (dashed line), normalised to
the data in the accepted area. The vertical lines indicate the cuts
applied to the data.}
\label{pic:lebensdauer_ppi_npi}
\end{figure}

In the next step, the decay time of $\Sigma^\pm$ was calculated
using the track length and the 3-momentum of the $\Sigma^\pm$. The
distributions are shown in figs. \ref{pic:lebensdauer-} and
\ref{pic:lebensdauer_ppi_npi} together with those of Monte-Carlo
simulated events. The residual background seen at large decay times
is due to secondary reactions in target and central drift chamber.
It is subtracted in the final background substraction.

The strong excess of events at small decay times  indicates
background from other reactions which accumulates at small times.
Due to the limited resolution this is expected if all charged
particles originate from the same production vertex. In order to
reduce this background, events with $t_{\Sigma^\pm} < 0.7  \cdot
10^{-10}\,$s were removed. For events with $\Sigma^{+}$ decay,
another cut was applied for large decay times. The cut
$t_{\Sigma^{+}} > 4.5\cdot 10^{-10}\,$s removed events in a region
where further background is visible.

The  data sets, obtained after the selection cuts, contained 4429
events from the reaction $\gamma p \rightarrow K^+\Sigma^-\pi^+ $
and 11267 events from $\gamma p \rightarrow K^+ \Sigma^+\pi^-$, with
5080 of the $\Sigma^+$ decaying into $n\pi^+$ and 6187 into
$p\pi^{o}$. Background contributions which were not removed by the
selection cuts described above were estimated by Monte-Carlo
simulations and finally subtracted (section \ref{sec:results}).

\section{Acceptance of the events}
\label{sec:acceptance}
The acceptance  was determined by  simulating events in the SAPHIR
setup for the reactions $\gamma p \rightarrow K^ +
\Sigma^{\pm}\pi^{\mp}$ according to phase space with propagation of
$\Sigma^{\pm}$ and subsequent decays $\Sigma^{-} \rightarrow
n\pi^{-}$ and $\Sigma^{+} \rightarrow p\pi^{0}$ or $\Sigma^{+}
\rightarrow n\pi^{+}$, respectively. Charged particles in the final
states were tracked through the drift chamber system taking into
account the magnetic field and multiple scattering in all materials.
Simulated events were processed like real events through the event
reconstruction and selection procedures. The total acceptance
accounted for the trigger efficiency of the data taking periods, the
event reconstruction efficiency and the data reduction according to
the event selection cuts. The mean acceptance was of the order of
$10\,\%$ for $\gamma p \rightarrow K^+\Sigma^{-}\pi^+$ and $2\,\%$
for $\gamma p \rightarrow K^+\Sigma^{+}\pi^-$. The acceptance of the
latter reaction was lower because it includes the efficiency of the
TOF measurements for both $\Sigma^+$ decay modes and, in addition,
the cut in the angular distribution of $\Sigma^+\rightarrow p\pi^0$
decays (see section \ref{selection}).

\section{Background from other reactions}
\label{sec:background}

Background  was estimated by generating events according to phase
space for the reactions listed in table \ref{tab:bg}. The events
were processed through reconstruction and selection criteria as real
events. The background event samples obtained were normalised
according to the photon flux.

The errors in the background estimate are dominated by a constant
value of 10\% due to the model dependence of the event simulation
and the uncertainty of the background cross sections. This error and
the statistical errors were added in quadrature. For  $\gamma p
\rightarrow K^ + \Sigma^{\pm}\pi^{\pm}$, $\Sigma^{\pm} \rightarrow n
\pi^{\mp}$, the reactions $\gamma p\rightarrow p\pi^+\pi^-\pi^0$ and
$\gamma p \rightarrow n\pi^+\pi^+\pi^-$ contribute on average with
about $10\%$, and all reactions together with $(13\pm1.4)\%$ to the
observed total cross section (see section\,\ref{sec:results}). For
$\gamma p \rightarrow K^ + \Sigma^{+}\pi^{-}$, $\Sigma^{+}
\rightarrow p \pi^{0}$, the reaction $\gamma p\rightarrow
p\pi^+\pi^-$ contributes on average with about $10\%$ and the total
background adds up to $(20\pm3)\%$ of the observed cross section.

\begin{table}[h!]
 \caption{Reactions and cross sections in the photon energy range
considered. The list includes $\gamma p \rightarrow K^+
\Sigma^{\pm}\pi^{\mp}$, since these reactions also contaminate each
other.} \centerline{
\begin{tabular}{lc}
Reaction &   $\sigma$ [$\mu b$]\\[1mm]\hline
$\gamma p \rightarrow n\pi^+\pi^+\pi^-$       &  2-10     \\
$\gamma p \rightarrow p\pi^+\pi^-\pi^0$       &  5-25    \\
$\gamma p \rightarrow p\pi^+\pi^-$            &  58-30    \\[1mm]
$\gamma p \rightarrow K^+\Sigma^-\pi^+$       &  0-0.3   \\
$\gamma p \rightarrow K^+\Sigma^+\pi^-$       &  0-0.7    \\[1mm]
$\gamma p \rightarrow K^+ \Lambda \pi^0$      &  $< 1$     \\
$\gamma p \rightarrow K^0_{S} \Lambda \pi^+$  &  $< 1$     \\
$\gamma p \rightarrow K^0_{L} \Lambda \pi^+$  &  $< 1$    \\
$\gamma p \rightarrow K^0_{S} \Sigma^+ \pi^0$ & $< 1$     \\
$\gamma p \rightarrow K^+ \Lambda$            &  1.8-0.6   \\
$\gamma p \rightarrow K^+ \Sigma^0$           &  2.3-0.6     \\
$\gamma p \rightarrow K^+ \Lambda\pi^+\pi^-$  &  $\ll 1$     \\
$\gamma p \rightarrow K^+ \Sigma^0\pi^+\pi^-$ &  $\ll 1$  \\
\end{tabular}
\vspace{-8mm}}
\label{tab:bg}
\end{table}

\section{\boldmath $\Sigma\pi$ and $K\pi$ \unboldmath mass distributions}
\label{sec:invmass}

\begin{figure}[h]
\centerline{
\includegraphics[width=0.45\textwidth]{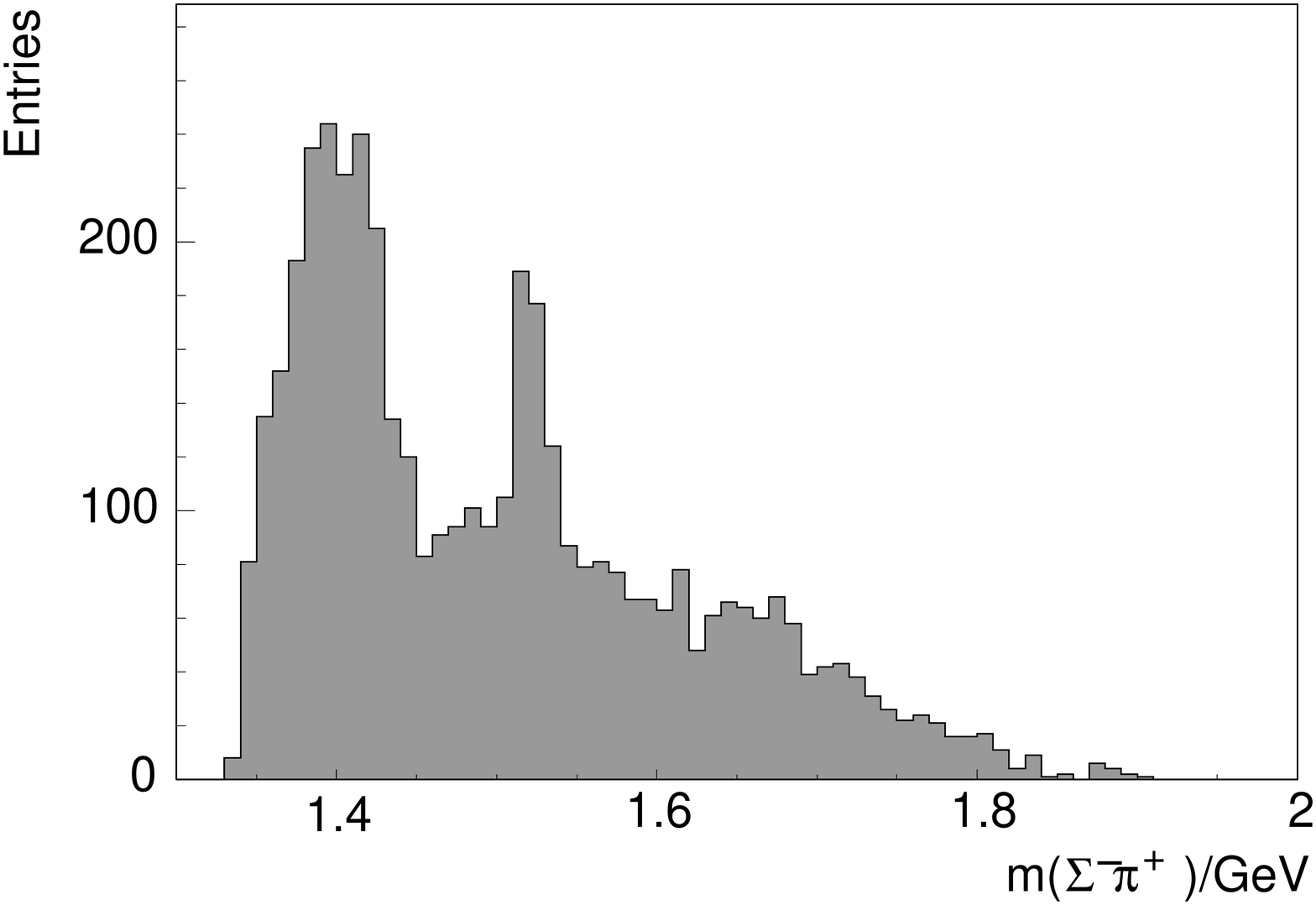}}
\caption{$\Sigma^-\pi^+$ invariant mass distribution.}
\label{pic:sigma-pi+}
\end{figure}

\begin{figure}[h]
\centerline{
\includegraphics[width=0.45\textwidth]{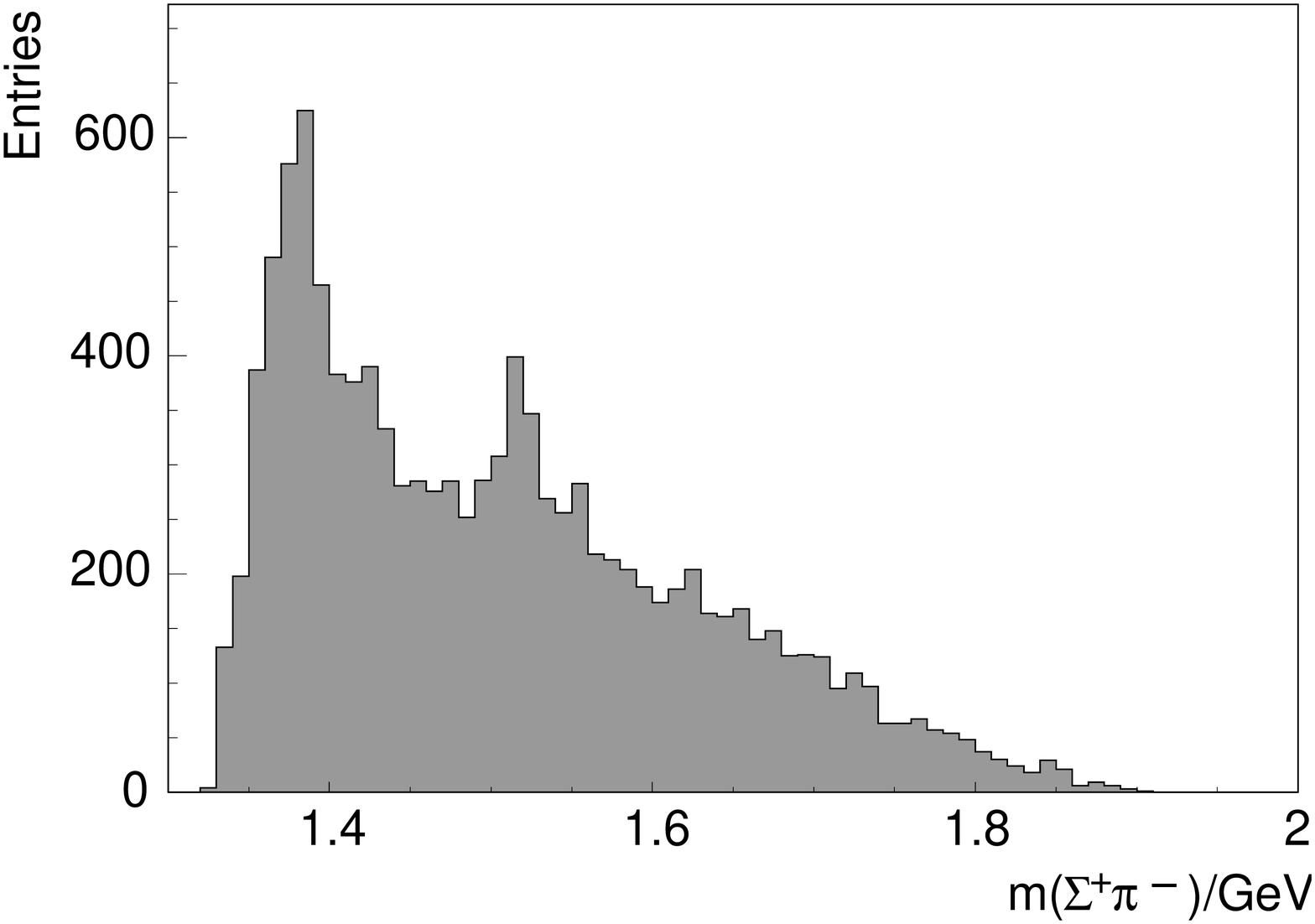}}
\caption{$\Sigma^+\pi^-$ invariant mass distribution.}
\label{pic:sigma+pi-}
\end{figure}

\begin{figure}[h]
\centerline{
\includegraphics[width=0.45\textwidth]{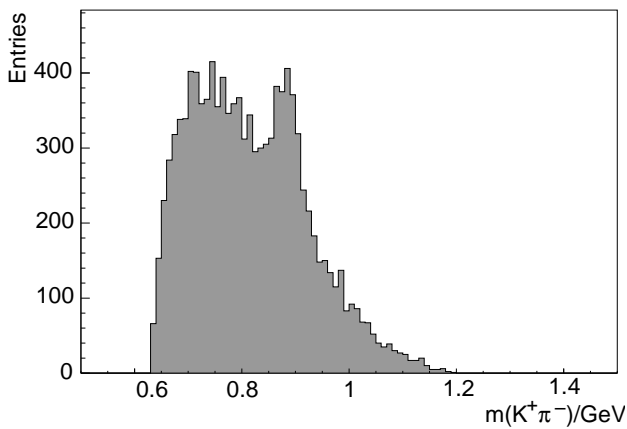}}
\caption{$K^+\pi^-$ invariant mass distribution.}
\label{pic:k+pi-}
\end{figure}

The invariant mass distributions for the $\Sigma\pi$ system  for the
reactions  $\gamma p \rightarrow K^ +\Sigma^{-}\pi^{+}$ and $\gamma
p \rightarrow K^{+} \Sigma^{+}\pi^{-}$ are shown in figs.
\ref{pic:sigma-pi+} and \ref{pic:sigma+pi-}. Both, $\Sigma^+\pi^-$
and $\Sigma^-\pi^+$ distributions show a  peak structure in the mass
range of $\Sigma(1385)$ and $\Lambda(1405)$ and another  pronounced
peak in the mass range of $\Lambda(1520)$. Figure \ref{pic:k+pi-}
shows the $K^+\pi^-$ mass distribution for events assigned to the
reaction $\gamma p \rightarrow K^{+} \Sigma^{+}\pi^{-}$. The peak at
890 MeV indicates $K^{*0}$ production.

From the observed resonance peaks it can be concluded that
substantial parts of both reaction cross sections are due to
intermediate two-body resonant states. The $\Lambda(1520)$
production was studied in detail and is presented in a separate
paper \cite{Lambda1520}.

\section{Reaction cross sections}
\label{sec:results}
Cross sections were determined as a function of the photon energy for both
reactions, in case of $\gamma p\rightarrow K^{+}\Sigma^{+}\pi^{-}$ separately
for both  $\Sigma^{+}$ decay modes.

\begin{figure}[ht]
\centerline{\includegraphics[width=0.45\textwidth]{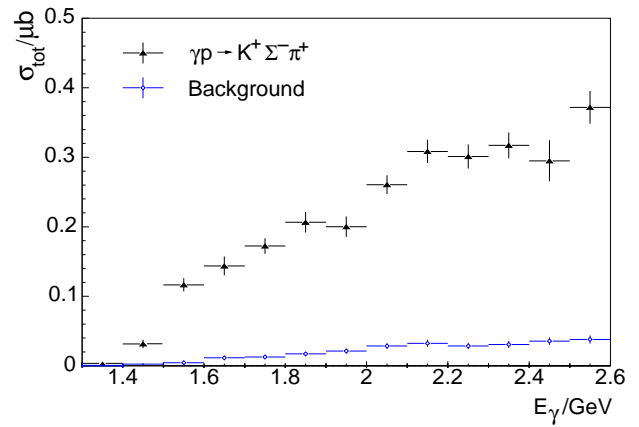}}
\caption{Excitation function before background subtraction for
$\gamma p \rightarrow K^{+}\Sigma^{-}\pi^{+}$ and background from
other reactions.}
\label{pic:wqsigma-mitbg}
\end{figure}

\begin{figure}[ht!]
\centerline{\includegraphics[width=0.45\textwidth]{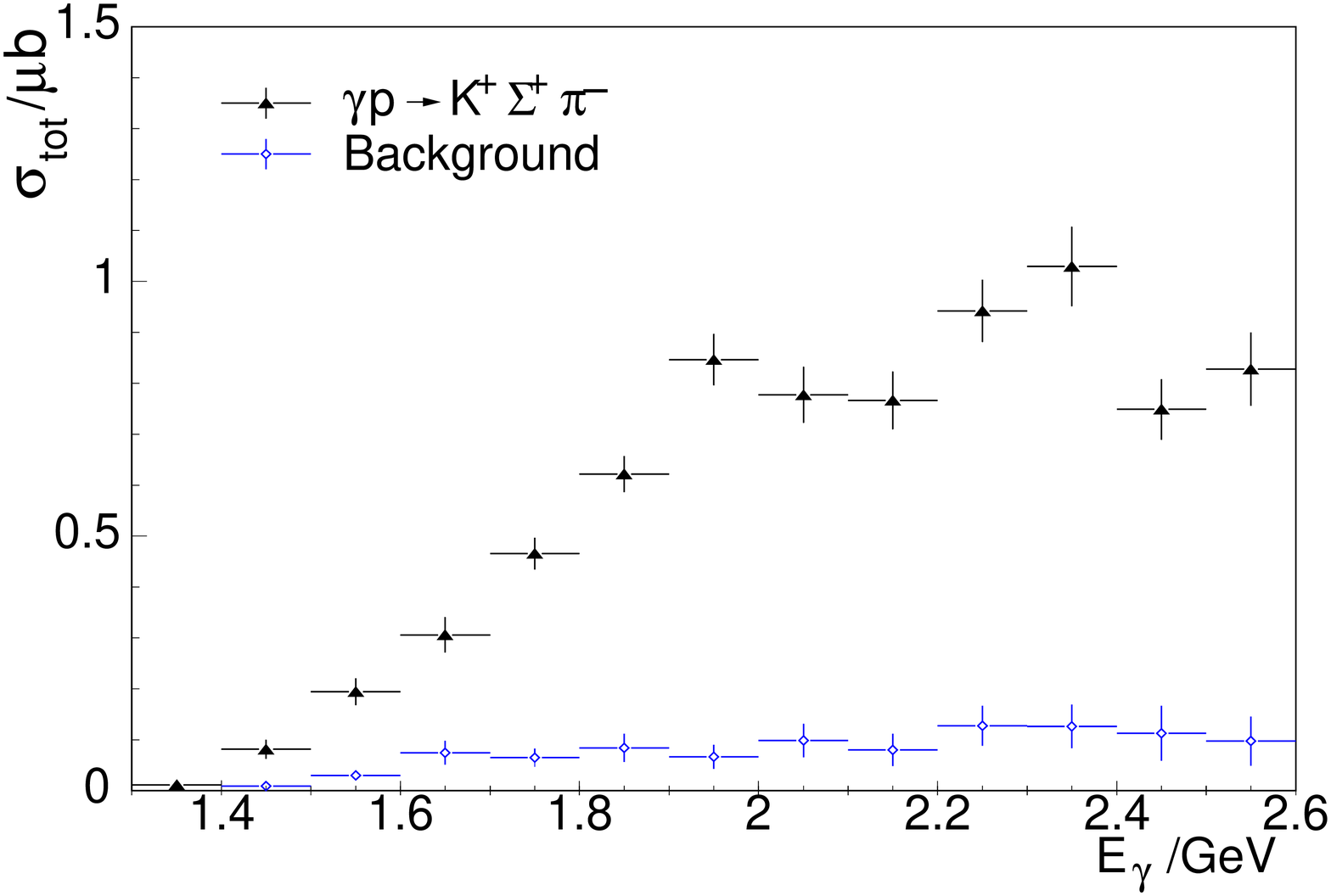}}
\caption{Excitation function before background subtraction for the
reaction $\gamma p \rightarrow K^{+}\Sigma^{+}\pi^{-}$ with decay
$\Sigma^{+} \rightarrow n  \pi^{+}$
 and background from other reactions.}
\label{pic:wqsigma+npi-mitbg}
\end{figure}

\begin{figure}[ht!]
\centerline{\includegraphics[width=0.45\textwidth]{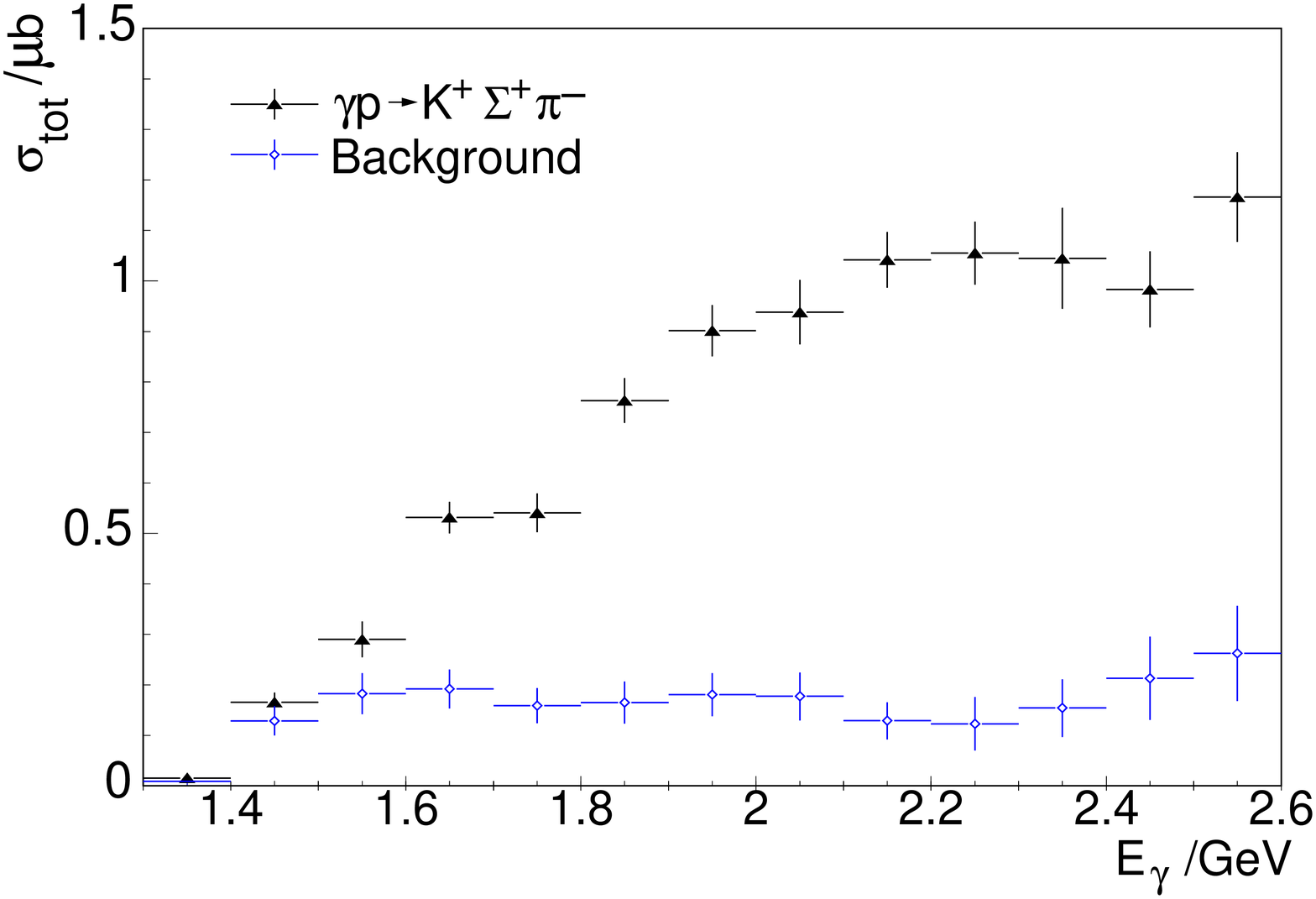}}
\caption{Excitation function before background subtraction for the
reaction $\gamma p \rightarrow K^{+}\Sigma^{+}\pi^{-}$ with decay
$\Sigma^{+} \rightarrow p\pi^{0}$
 and background from other reactions.}
\label{pic:wqsigma+ppi-mitbg}
\end{figure}

Figures \ref{pic:wqsigma-mitbg}, \ref{pic:wqsigma+npi-mitbg} and
\ref{pic:wqsigma+ppi-mitbg}
show the excitation function before background subtraction
with statistical errors
together with the total
background contributions according to section \ref{sec:background}.

\begin{figure}[ht!]
\centerline{\includegraphics[width=0.45\textwidth]{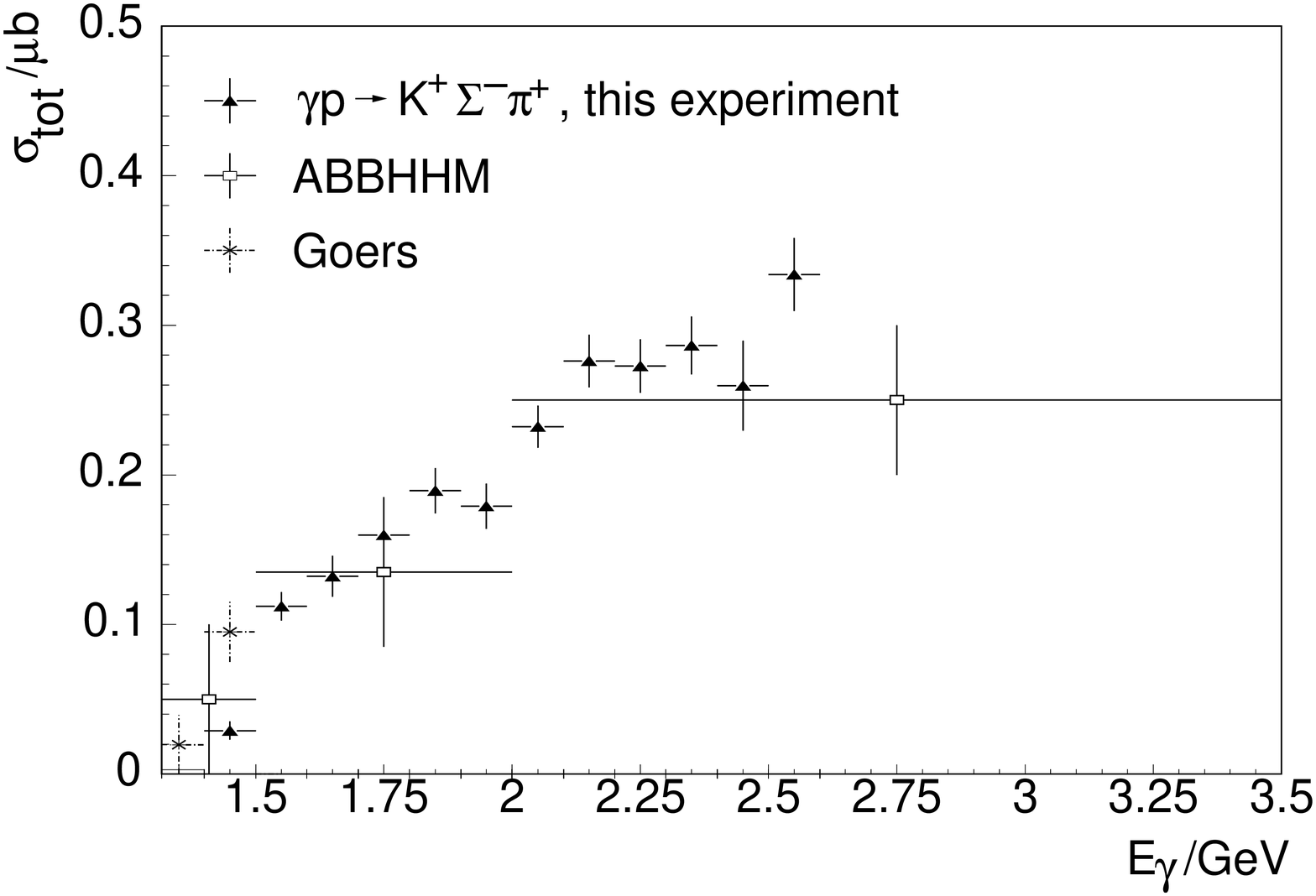}}
\caption{Cross section of the reaction $\gamma p \rightarrow
K^{+}\Sigma^{-}\pi^{+}$
as a function of the photon energy after
  subtraction of background from
  other reactions in comparison to previous measurements \cite{ABBHHM,SGoers}.
}
\label{pic:wq-ohnebg}
\end{figure}

\begin{figure}[ht!]
\centerline{\includegraphics[width=0.45\textwidth]{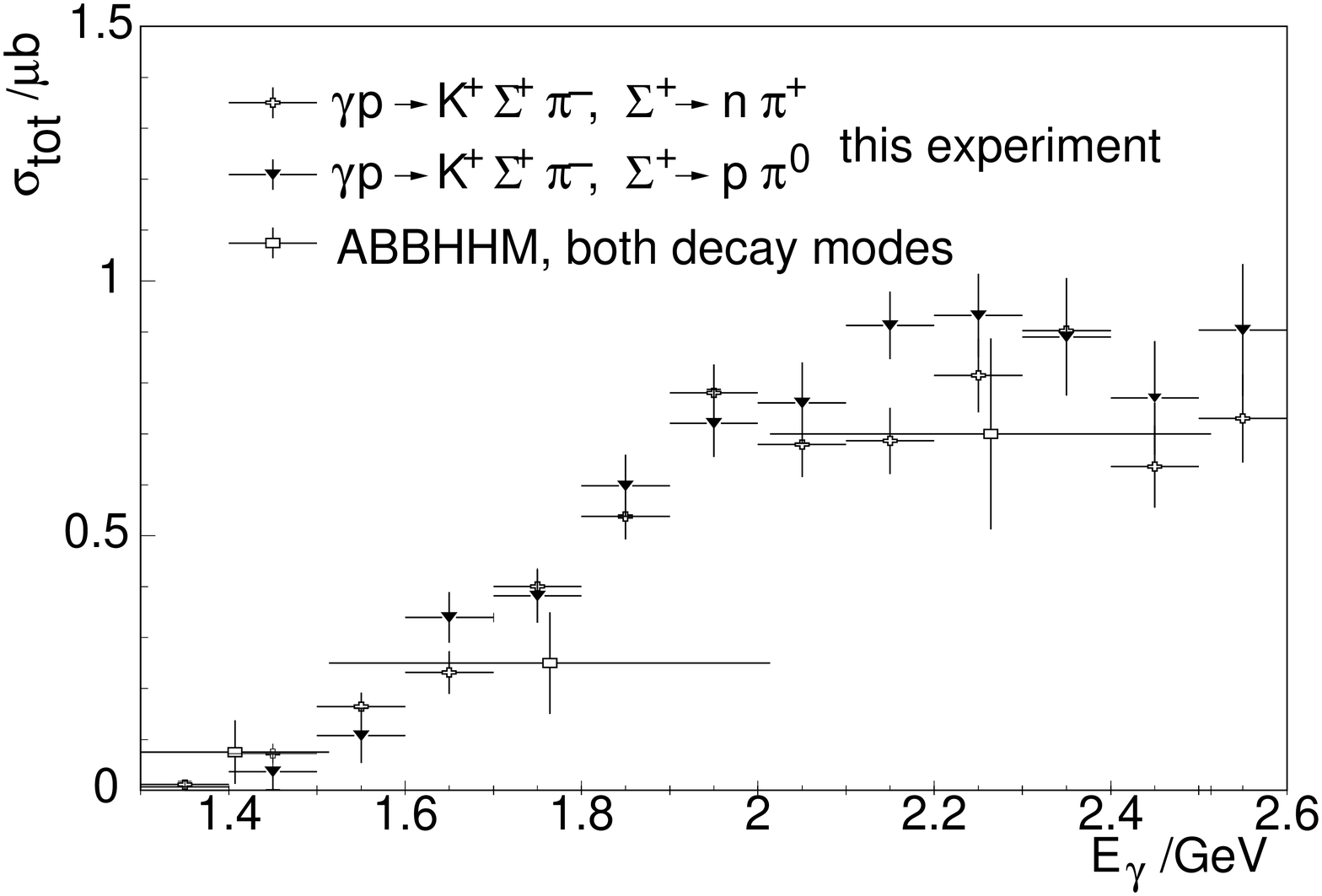}}
\caption{Cross section of the reaction $\gamma p \rightarrow K^{+}\Sigma^{+}\pi^{-}$
as a function of the photon energy after background subtraction from
other reactions in comparison to previous measurements \cite{ABBHHM}. For this experiment
the cross sections from both $\Sigma^+$ decay modes are given separately. }
\label{pic:wq+ohnebg}
\end{figure}

The final reaction cross sections were obtained by subtracting  the
accumulated background cross sections bin-by-bin. They are shown in
figs. \ref{pic:wq-ohnebg} and \ref{pic:wq+ohnebg}. The errors were
calculated by quadratic addition. Cross sections and errors are
quoted in tables \ref{tab:tot_wq1}, \ref{tab:tot_wq2}, and
\ref{tab:tot_wq3}, respectively.

\section{Summary}
\label{sec:summary}
The cross sections of the reactions $\gamma p\rightarrow K^+
\Sigma^{\pm}\pi^{\mp}$ were measured in the photon energy range from
threshold  to 2.6 GeV. They rise monotonously up to values of about
0.3 $\mu$b for $K^{+}\Sigma^{-}\pi^{+}$ and about 0.8 $\mu $b for
$K^{+}\Sigma^{+} \pi^{-}$. Regarding the hitherto existing data an
evident improvement concerning the energy resolution and the total
errors is achieved. No indications are found for narrow structures
in the total cross sections, nor strong threshold enhancements as
seen, e.\,g., in $\gamma p \rightarrow p\eta$ \cite{Krusche:1995nv},
$\gamma p \rightarrow \Lambda K^+$ \cite{Glan03b,bleckmann,clas}, or
$\gamma p \rightarrow \omega p$ \cite{barthomega}. The $\Sigma\pi$
and $K\pi$ mass spectra show pronounced peak structures, indicating
that a substantial part of the cross sections is due to two-body
intermediate  states. The $\Lambda(1520)$ intermediate state is
investigated in a separate paper.

\begin{table}
\caption{\label{tab:tot_wq1}Total cross sections of the reaction
$\gamma p \rightarrow K^{+}\Sigma^{-}\pi^{+}$ for 13 bins of
$E_{\gamma}$, obtained after background subtraction.} \centerline{
\begin{tabular}{lll}
\hline
$E_\gamma$ [GeV]   & $\sigma_{tot}$ [$\mu b$] & $\delta\sigma$ [$\mu b$]  \\  \hline
$1.300\,-\,1.400$  & 0.0028        & $\pm$0.0026  \\
$1.400\,-\,1.500$  & 0.0290        & $\pm$0.0061  \\
$1.500\,-\,1.600$  & 0.1121        & $\pm$0.0095  \\
$1.600\,-\,1.700$  &  0.1322       & $\pm$0.0137  \\
$1.700\,-\,1.800$  &  0.1597       & $\pm$0.0113  \\
$1.800\,-\,1.900$  &  0.1894       & $\pm$0.0152  \\
$1.900\,-\,1.000$  &  0.1790       & $\pm$0.0152  \\
$2.000\,-\,1.100$  &  0.2322       & $\pm$0.0141  \\
$2.100\,-\,2.200$  &  0.2761       & $\pm$0.0177  \\
$2.200\,-\,2.300$  &  0.2727       & $\pm$0.0179  \\
$2.300\,-\,2.400$  &  0.2865       & $\pm$0.0194  \\
$2.400\,-\,2.500$  &  0.2595       & $\pm$0.0300  \\
$2.500\,-\,2.600$  &  0.3339       & $\pm$0.0243  \\ \hline
\end{tabular} \vspace{5mm}}
\caption{\label{tab:tot_wq2}Total cross sections of the reaction
$\gamma p \rightarrow K^{+}\Sigma^{+}\pi^{-}$ with $\Sigma^{+}
\rightarrow n \pi^{+}$
 for 13 bins of $E_{\gamma}$,
obtained after background subtraction.}
\centerline{
\begin{tabular}{lll}
\hline
$E_\gamma$ [GeV]   & $\sigma_{tot}$ [$\mu b$] & $\delta\sigma$ [$\mu b$]  \\  \hline
$1.300\,-\,1.400$  & 0.0116       & $\pm$0.0076  \\
$1.400\,-\,1.500$  & 0.0720       & $\pm$0.0191  \\
$1.500\,-\,1.600$  & 0.1644       & $\pm$0.0275  \\
$1.600\,-\,1.700$  & 0.2314       & $\pm$0.0419  \\
$1.700\,-\,1.800$  & 0.4005       & $\pm$0.0359  \\
$1.800\,-\,1.900$  & 0.5377       & $\pm$0.0448  \\
$1.900\,-\,1.000$  & 0.7802       & $\pm$0.0559  \\
$2.000\,-\,1.100$  & 0.6789       & $\pm$0.0641  \\
$2.100\,-\,2.200$  & 0.6861       & $\pm$0.0650  \\
$2.200\,-\,2.300$  & 0.8147       & $\pm$0.0726  \\
$2.300\,-\,2.400$  & 0.9029       & $\pm$0.0894  \\
$2.400\,-\,2.500$  & 0.6359       & $\pm$0.0806  \\
$2.500\,-\,2.600$  & 0.7302       & $\pm$0.0868  \\ \hline
\end{tabular}
\vspace{5mm}}
\caption{\label{tab:tot_wq3} Total cross sections of the reaction
$\gamma p \rightarrow K^{+}\Sigma^{+}\pi^{-}$ with $\Sigma^{+}
\rightarrow p \pi^{0}$ for
 13 bins of $E_{\gamma}$,
obtained after background subtraction.} \centerline{
\begin{tabular}{lll}
\hline
$E_\gamma$ [GeV]   & $\sigma_{tot}$ [$\mu b$] & $\delta\sigma$  \\  \hline
$1.300\,-\,1.400$  & 0.0067        & $\pm$0.0113  \\
$1.400\,-\,1.500$  & 0.0367        & $\pm$0.0347  \\
$1.500\,-\,1.600$  & 0.1075        & $\pm$0.0538  \\
$1.600\,-\,1.700$  &  0.3395       & $\pm$0.0496  \\
$1.700\,-\,1.800$  &  0.3817       & $\pm$0.0519  \\
$1.800\,-\,1.900$  &  0.5983       & $\pm$0.0605  \\
$1.900\,-\,1.000$  &  0.7209       & $\pm$0.0662  \\
$2.000\,-\,1.100$  &  0.7605       & $\pm$0.0794  \\
$2.100\,-\,2.200$  &  0.9128       & $\pm$0.0662  \\
$2.200\,-\,2.300$  &  0.9323       & $\pm$0.0816  \\
$2.300\,-\,2.400$  &  0.8905       & $\pm$0.1151  \\
$2.400\,-\,2.500$  &  0.7702       & $\pm$0.1115  \\
$2.500\,-\,2.600$  &  0.9037       & $\pm$0.  \\ \hline
\end{tabular}
\vspace{-1.0cm}}
\end{table}

\section{Acknowledgements}
\label{sec:acknoledgements}

We would like to thank the technical staff of the ELSA machine group
for their invaluable contributions to the experiment. We gratefully
acknowledge the support by the Deutsche Forschungsgemeinschaft in
the framework of the Schwerpunktprogramm ``Investigation of the
hadronic structure of nucleons and nuclei with electromagnetic
probes'' (SPP 1034 KL 980/2-3) and the Sonderforschungsbereich
SFB/TR16 (``Subnuclear Structure of Matter'').



\begin{thebibliography}{}

\bibitem{klempt}
E.\ Klempt, J.-M.\ Richard, Rev.\ Mod.\ Phys.\ 82, 1095–1153 (2010).
%
\bibitem{Glan03b}
K.-H.\ Glander {\it et al.}, Eur.\ Phys.\ J.\ A 19, 251 (2004).
%
\bibitem{Goers99}
S.\ Goers {\it et al.}, Phys.\ Lett.\ B 464, 331 (1999).
%
\bibitem{Lawall05}
R.\ Lawall {\it et al.}, Eur.\ Phys.\ J.\ A 24, 275 (2005).
%
\bibitem{wu}
C.\ Wu {\it et al.}: Eur.\ Phys.\ J.\ A 23 (2) (2005).
%
\bibitem{barthomega}
J.\ Barth {\it et al.}: Eur.\ Phys.\ J.\ A 18 117-127 (2003).
%
\bibitem{barthphi}
J.\ Barth {\it et al.}:   Eur.\ Phys.\ J.\ A 17 2, 269-274 (2003).
%
\bibitem{etaprime}
R.\ Pl\"{o}tzke {\it et al.}: Phys.\ Lett.\ B 444 555-562 (1998).
%
\bibitem{Schwille}
W.\ J.\ Schwille {\it et al.}, Nucl.\ Instr.\ Meth. A 344, 470 (1994).
%
\bibitem{Hillert}
W.\ Hillert, Eur.\ Phys.\ J.\ A28, 139 (2006).
%
\bibitem{Lambda1520}
F.\ W.\ Wieland {\it et al.}, preceding paper.
%
\bibitem{ABBHHM}
R.\ Erbe {\it et al.} (ABBHHM), Physical Review 188, 2060 (1969).
%
\bibitem{SGoers}
S.\ Goers, doctoral thesis, Bonn University (1999), BONN-IR-1999-09.
%
\bibitem{Krusche:1995nv}
B.\ Krusche {\it et al.}, Phys.\ Rev.\ Lett.\ {\bf 74}, 3736 (1995).
%
\bibitem{bleckmann}
A.\ Bleckmann {\it et al.}, Z.\ Phys.\ {\bf 239}, 1 (1970).
%
\bibitem{clas}
R.\ Bradford {\it et al.}, Phys.\ Rev.\ C 73, 035202 (2006).
%
\end{thebibliography}
\end{document}